\documentclass[aip, jcp, reprint]{revtex4-1}

\usepackage{amsmath, amsfonts}
\usepackage{graphicx, url}
\usepackage{natbib}
\usepackage{multirow, rotating}
\usepackage{hyperref}
\hypersetup{colorlinks = true, linktocpage = true, bookmarksopen = true, linkcolor = blue, urlcolor=blue, citecolor = blue, allcolors=blue}
\usepackage{color}
\def\cbl{\color{black}}
\def\cb{\color{black}}

\begin{document}
\title{\cbl New homogenization approaches for stochastic transport through heterogeneous media \cb}
\author{Elliot J. Carr}
\author{Matthew J. Simpson}
\affiliation{School of Mathematical Sciences, Queensland University of Technology, Brisbane, Australia.}

\begin{abstract}
The diffusion of molecules in complex intracellular environments can be strongly influenced by spatial heterogeneity and stochasticity. A key challenge when modelling such processes using stochastic random walk frameworks is that negative jump coefficients can arise when transport operators are discretized on heterogeneous domains.  Often this is dealt with through homogenization approximations by replacing the heterogeneous medium with an \textit{effective} homogeneous medium.  In this work, we present a new class of homogenization approximations by considering a stochastic diffusive transport model on a one-dimensional domain containing an arbitrary number of layers with different jump rates.  We derive closed form solutions for the $k$th moment of particle lifetime, carefully explaining how to deal with the internal interfaces between layers.  These general tools allow us to derive simple formulae for the effective transport coefficients, leading to significant generalisations of previous homogenization approaches. Here, we find that different jump rates in the layers gives rise to a net bias, leading to a non-zero advection, for the entire homogenized system. Example calculations show that our generalized approach can lead to very different outcomes than traditional approaches, thereby having the potential to significantly affect simulation studies that use homogenization approximations.
\end{abstract}
\maketitle

\section{Introduction}
The motion of cells in heterogeneous tissues and the diffusion of molecules in complex intracellular environments is strongly influenced by spatial heterogeneity~\cite{Saxton1994,Swanson2000,Isaacson2006,Lepzelter2012,Carr2018b}.  Modelling transport through heterogeneous materials is far more challenging than modelling transport through homogeneous materials.  Mathematical models of transport through homogeneous materials can often be solved analytically, whereas mathematical models of transport through heterogeneous materials often require repeated numerical calculations, thereby providing less general insight.

The motion of cells and molecules in biological tissues is often modelled using stochastic approaches~\cite{Codling2008,Allen2010,Isaacson2013}.  A key challenge is that negative jump coefficients can arise when transport operators are discretized on domains characterized by heterogeneous transport coefficients~\cite{Lotstedt2015,Meinecke2016a,Meinecke2016b,Meinecke2017,Engblom2017}.  This problem can be circumvented through homogenization by replacing the heterogeneous medium with an \textit{effective} homogeneous medium~\cite{Lotstedt2015,Meinecke2016a,Meinecke2016b,Meinecke2017,Engblom2017}.  One of the challenges in using a homogenization approach is that there is a wide range of techniques and results available.  For continuum models, based on partial differential equations, transport equations can either be homogenized using volume averaging or asymptotic expansions~\cite{Davit2013}.  For stochastic models of transport through heterogeneous media, there are also several homogenization approximations available for different applications~\cite{Derrida1982,Berezhkovskii2003,Kalnin2013,Kalnin2015}.

We present a new approach for homogenization by considering a stochastic, lattice-based transport model on an interval $[0,L]$ partitioned into $m$ layers, $(x_{i-1},x_{i})$ for $i = 1,\hdots, m$, where $0 = x_{0} < x_{1} < x_{2} <\hdots < x_{m-1} < x_{m} = L$ and $x_{i}$ denotes the location of the interface between layer $i$ and layer $i + 1$ for $i = 1,\hdots, m-1$~\cite{Carr2017b}. An agent, initialized at some location, undergoes a random walk with probability $P_i$ of taking one step in the positive $x$ direction and probability $P_i$ of taking one step in the negative $x$ direction during each time step when located in the interior of layer $i$.  A key tool used to describe this kind of model in a homogeneous setting is the mean particle lifetime or mean first exit time~\cite{Lotstedt2015,Redner01,Vaccario2015}.  Here we generalize this concept and provide exact, analytical tools that can be used to calculate the $k$th moment of particle lifetime in an arbitrarily heterogeneous medium~\cite{Ellery2012a,Ellery2012b,Carr2017a,Simpson2013,Carr2018a,Gordon2013}.  The new method is very powerful as it leads to exact closed form expressions  for any moment of particle lifetime, $k=1,2,3,\ldots$, in an arbitrary system composed of any number of layers, $m=1,2,3\ldots$.  With these tools we derive simple formulae for the effective transport coefficients, leading to significant generalisations of previous results.  Example calculations show that traditional approximations can lead to very different outcomes that may not capture the appropriate underlying physics of interest.

\section{Stochastic transport model}
\label{sec:stochastic_transport_model}

\begin{figure*}[t]
\centering
\includegraphics[width=1.0\textwidth]{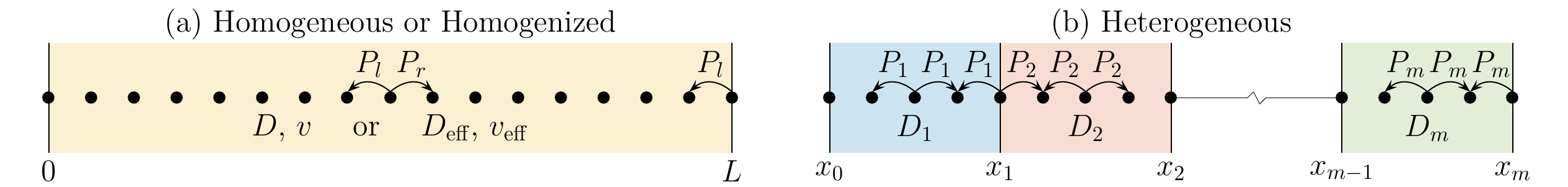}
\caption{(a) Biased random walk in a homogeneous environment on the interval $[0,L]$. A particle takes one step in the positive $x$ direction with probability $P_{r}$ and one step in the negative $x$ direction with probability $P_{l}$. The continuum limit of the discrete stochastic model is characterized by two transport properties, diffusivity and drift, represented by $D$ and $v$ for the homogeneous model and $D_{\mathrm{eff}}$ and $v_{\mathrm{eff}}$ for the homogenized model (b) Unbiased random walk in a heterogeneous environment, where the interval $[x_0,x_m] = [0,L]$ is partitioned into $m$ layers. A particle in the $i$th layer takes one step in the positive $x$ direction with probability $P_{i}$ and one step in the negative $x$ direction with probability $P_{i}$. The continuum limit of the discrete stochastic model is characterized by an individual diffusivity $D_{i}$ in the $i$th layer.}
\label{fig:1}
\end{figure*}

We consider a random walk in both a homogeneous environment, shown in Fig \ref{fig:1}(a), and a heterogeneous environment containing two or more layered homogeneous materials with different material properties, shown in Fig \ref{fig:1}(b). In both cases, the one-dimensional Cartesian geometry $[0,L]$ is discretized to give $N$ lattice sites with uniform spacing, $\Delta = L/(N-1) > 0$. Sites are indexed so that site $j$ has position $x = (j-1)\Delta$ for all $j = 1,\hdots,N$. For the heterogeneous environment, we choose $N$ to ensure that a site is located at each interface ($x = x_{i}$, $i = 1,\hdots, m$) as shown in Fig \ref{fig:1}(b). A particle is placed on the lattice, and during each time step of duration $\tau > 0$, the particle undergoes a nearest-neighbour random walk. Any walker reaching the left boundary, $j=1$, is removed. In contrast, the right boundary, $j=N$, is reflecting.

Let $\mathbb{E}(T_{j})$ be the mean time for a particle released at site $j$ to exit the system, and $\mathbb{E}(T^{(k)}_{j})$ to be the $k$th moment of the lifetime distribution of an ensemble of particles released at site $j$.  The family of moments, $k=1,2,3, \ldots$, can be written as
\begin{equation}\label{eq:Momentdefinition}
\mathbb{E}(T^{(k)}_{j}) = \sum_{l=0}^{\infty} t^k \, \mathbb{P}(T_{j} = t),
\end{equation}
where $t = l\tau$ is time, $\mathbb{E}(T_{j}^{(0)}) = 1$ and $\mathbb{P}(T_{j} = t)$ is the probability that an agent released at site $j$ exits the system at time $t$.  Our aim is to arrive at a discrete relationship involving $\mathbb{E}(T^{(k)}_{j})$ for both the homogeneous and heterogeneous environments, which can then be converted into a family of boundary value problems for which we can derive exact closed-form solutions.

\textit{Homogeneous environment.}\ We consider both a biased and an unbiased random walk in the homogeneous environment, Fig \ref{fig:1}(a). A particle currently located at a lattice site in the interior of the medium has three possible outcomes during the next time step: (i) the particle takes one step in the positive $x$ direction with probability $P_{r}\in(0,1)$; (ii) the particle takes one step in the negative $x$ direction with probability $P_{l}\in(0,1-P_{r}]$; or (iii) the particle remains at the same location with probability $1-P_{l}-P_{r}$. Considering a particle initially released at lattice site $j$ and conditioning on the three possible outcomes during the first step of the stochastic process, we have
\begin{gather}\nonumber
\mathbb{P}(T_{j} = t) = \mathbb{P}(T_{j-1} = t-\tau)P_{l} + \mathbb{P}(T_{j+1} = t-\tau)P_{r}\\\label{eq:conditionalstatement} + \mathbb{P}(T_{j} = t-\tau)(1-P_{r}-P_{l}).
\end{gather}
If we first consider $k=1$, substituting Eq (\ref{eq:conditionalstatement}) into Eq (\ref{eq:Momentdefinition}), re-writing $t$ as $\left[ (t-\tau) + \tau \right]$, and rearranging, we obtain
\begin{equation}
P_{l}\left[\mathbb{E}(T^{(1)}_{j-1}) - \mathbb{E}(T^{(1)}_{j})\right] + P_{r}\left[\mathbb{E}(T^{(1)}_{j+1}) - \mathbb{E}(T^{(1)}_{j})\right] = -\tau,
\end{equation}
which can be thought of as a discrete conservation statement for $\mathbb{E}(T_{j}^{(1)})$. \cbl Repeating the process for $k=2,3$ is sufficient to observe a pattern that can be formalized by induction to give\cb
\begin{gather}\nonumber
P_{l}\left[\mathbb{E}(T^{(k)}_{j-1}) - \mathbb{E}(T^{(k)}_{j})\right] + P_{r}\left[\mathbb{E}(T^{(k)}_{j+1}) - \mathbb{E}(T^{(k)}_{j})\right]\\\label{eq:generaldiscretemoment} = \sum_{l=0}^{k-1}{k \choose l}(-\tau)^{k-l}\mathbb{E}\left(T^{(l)}_{j}\right),
\end{gather}
for all $k = 1,2,\hdots$. To convert this family of discrete conservation statements into a continuum model we identify the discrete moment $\mathbb{E}(T^{(k)}_{j})$ with a smooth continuous function $\mathbb{M}_k(x)$. Expressing $\mathbb{E}(T^{(k)}_{j \pm 1})$ in Eq (\ref{eq:generaldiscretemoment}) in terms of appropriate Taylor series expansions
\begin{align*}
\mathbb{E}(T_{j\pm 1}^{(k)}) &\equiv \mathbb{M}_k(x) \pm \Delta \dfrac{\textrm{d} \mathbb{M}_k(x)}{\textrm{d} x} + \dfrac{\Delta^2}{2} \dfrac{\textrm{d}^2\mathbb{M}_k(x)}{\textrm{d} x^2} + \mathcal{O}(\Delta^3),
\end{align*}
assuming $P_{r}-P_{l} = \mathcal{O}(\Delta)$~\cite{Codling2008} and considering the limit as $\Delta \to 0$ and $\tau \to 0$ jointly in Eq (\ref{eq:generaldiscretemoment}) with the ratio $\Delta^2/\tau$ held finite leads to the differential equation:
\begin{gather}\label{eq:Mk_ode_biased}
D\dfrac{\textrm{d}^2\mathbb{M}_k(x)}{\textrm{d} x^2} - v\dfrac{\textrm{d}\mathbb{M}_k(x)}{\textrm{d} x} = -k\mathbb{M}_{k-1}(x),
\end{gather}
for all $k=1,2,3,\ldots$, with $\mathbb{M}_0(x)=1$. The appropriate boundary conditions are:
\begin{gather}\label{eq:Mk_bcs_biased}
\mathbb{M}_{k}(0) = 0,\quad \frac{\text{d}\mathbb{M}_{k}(L)}{\text{d}x} = 0,
\end{gather}
with the diffusion and drift coefficients, $D$ and $v$, defined as~\cite{Codling2008,Hughes1995}:
\begin{equation}
\label{eq:D_v}
D = \frac{(P_{r}+P_{l})\Delta^{2}}{2\tau},\quad v = \frac{(P_{l}-P_{r})\Delta}{\tau}.
\end{equation}
Note that $v < 0$ if $P_{r}>P_{l}$ so the $k$th moment is advected in the opposite direction to that of a particle. For an unbiased random walk, where $P_{l} = P_{r}$ and $v = 0$, we use the notation $M_{k}(x)$ to denote the continuous representation of $\mathbb{E}(T^{(k)}_{j})$ to distinguish it from $\mathbb{M}_k(x)$. Here, $M_{k}(x)$ satisfies the boundary value problem:
\begin{gather}\label{eq:Mk_ode_unbiased}
D  \dfrac{\textrm{d}^2 M_k(x)}{\textrm{d} x^2} = -k M_{k-1}(x),\\
\label{eq:Mk_bcs_unbiased}
M_{k}(0) = 0,\quad \frac{\text{d}M_{k}(L)}{\text{d}x} = 0,
\end{gather}
for all $k=1,2,3,\ldots$, with $M_0(x)=1$. \cbl The solution of Eqs (\ref{eq:Mk_ode_unbiased})--(\ref{eq:Mk_bcs_unbiased}), gives closed-form expressions for the $k$th moment of particle lifetime through studying the solutions of a family of coupled elliptic boundary value problems.  Alternatively, the same result can be obtained by solving the associated parabolic partial differential equation for the expected position of the particle in Laplace space~\cite{Redner01}.  Working in Laplace transform space can sometimes provide advantages by simplifying the algebra for these types of problems~\cite{Ellery2013}.  However, in this work, we always obtain closed-form expressions for the $k$th moment of particle lifetime without needing to work in Laplace transform space. \cb

\textit{Heterogeneous environment.}\ We consider an unbiased random walk in the heterogeneous environment, Fig \ref{fig:1}(b). A particle currently located at a lattice site in the interior of the $i$th layer has three possible outcomes during the next time step: (i) the particle takes one step in the positive $x$ direction with probability $P_{i}\in(0,0.5]$; (ii) the particle takes one step in the negative $x$ direction with probability $P_{i}$; or (iii) the particle remains at the same location with probability $1-2P_{i}$. Here, the probability $P_{i}$ is indexed by the layer number to signify that it may vary across layers. For this configuration, the $k$th moment of particle lifetime satisfies the discrete relationship given in Eq (\ref{eq:generaldiscretemoment}) with $P_{r} = P_{l} = P_{i}$:
\begin{gather}\nonumber
P_{i}\left[\mathbb{E}(T^{(k)}_{j-1}) + \mathbb{E}(T^{(k)}_{j+1}) - 2\mathbb{E}(T^{(k)}_{j})\right]\\\label{eq:heterogeneous_generaldiscretemoment} = \sum_{l=0}^{k-1}{k \choose l}(-\tau)^{k-l}\mathbb{E}(T^{(l)}_{j}).
\end{gather}
To convert this family of discrete conservation statements into a continuum model we identify $\mathbb{E}(T_{j}^{(k)})$ with a smooth continuous function $M_{k}^{(i)}(x)$ in each layer $i = 1,\hdots,m$, where $x_{i-1}<x<x_{i}$. Expressing $\mathbb{E}(T^{(k)}_{j \pm 1})$ in Eq (\ref{eq:heterogeneous_generaldiscretemoment}) in terms of appropriate Taylor series expansions:
\begin{gather*}
\mathbb{E}(T_{j\pm 1}^{(k)}) \equiv  M_{k}^{(i)}(x) \pm \Delta \dfrac{\textrm{d} M_{k}^{(i)}(x)}{\textrm{d} x} + \dfrac{\Delta^2}{2} \dfrac{\textrm{d}^2 M_{k}^{(i)}(x)}{\textrm{d} x^2}\\ + \mathcal{O}(\Delta^3),
\end{gather*}
yields:
\begin{gather}
\label{eq:order3}
P_{i}\Delta^{2} \dfrac{\textrm{d}^2 M_{k}^{(i)}}{\textrm{d} x^2} + \mathcal{O}(\Delta^3) = \sum_{l=0}^{k-1}{k \choose l}(-\tau)^{k-l}M_{l}^{(i)}(x).
\end{gather}
Dividing Eq (\ref{eq:order3}) by $\tau$ and considering the limit as $\Delta\rightarrow 0$ and $\tau\rightarrow 0$ jointly with the ratio $\Delta^{2}/\tau$ held finite leads to family of  differential equations:
\begin{gather}\label{eq:Mk_hetero_ode}
D_{i}\dfrac{\textrm{d}^2 M_{k}^{(i)}}{\textrm{d} x^2} =-kM_{k-1}^{(i)}(x), \quad x_{i-1}<x<x_{i},
\end{gather}
for $i = 1,\hdots,m$, where $D_{i} = P_{i}\Delta^{2}/\tau$ is the diffusivity associated with layer $i$. The appropriate boundary conditions are
\begin{gather}
\label{eq:Mk_hetero_bcs}
M_{k}^{(1)}(0) = 0,\quad \frac{\textrm{d}M_{k}^{(m)}(L)}{\textrm{d}x} = 0.
\end{gather}
To close the problem, Eqs (\ref{eq:Mk_hetero_ode})--(\ref{eq:Mk_hetero_bcs}) need to be paired with appropriate internal boundary conditions at the interfaces between adjacent layers ($x = x_{i}$, $i = 1,\hdots,m-1$). The first boundary condition is to assume continuity of the $k$th moment across the interface:
\begin{gather}
\label{eq:Mk_hetero_ic1}
M_{k}^{(i)}(x_{i}) = M_{k}^{(i+1)}(x_{i}),
\end{gather}
where $i = 1,\hdots,m-1$. The second boundary condition is derived by analysing the discrete relationship satisfied by the $k$th moment of particle lifetime at an arbitrary interface. We assume a particle currently located at the $i$th interface ($x = x_{i}$) has three possible outcomes during the next time step: (i) the particle takes one step in the positive $x$ direction with probability $P_{i+1}$; (ii) the particle takes one step in the negative $x$ direction with probability $P_{i}$; or (iii) the particle remains at the same location with probability $1-P_{i}-P_{i+1}$ (see Fig \ref{fig:1}(b)). Clearly, this requires $P_{i} + P_{i+1} \leq 1$ for all $i = 1,\hdots,m-1$. For this configuration, the $k$th moment of particle lifetime satisfies the discrete relationship given in Eq (\ref{eq:generaldiscretemoment}) with $P_{r} = P_{i+1}$ and $P_{l} = P_{i}$:
\begin{gather}\nonumber
P_{i}\left[\mathbb{E}(T^{(k)}_{j-1}) - \mathbb{E}(T^{(k)}_{j})\right] + P_{i+1}\left[\mathbb{E}(T^{(k)}_{j+1}) - \mathbb{E}(T^{(k)}_{j})\right]\\\label{eq:generaldiscretemoment_hetero} = \sum_{l=0}^{k-1}{k \choose l}(-\tau)^{k-l}\mathbb{E}(T^{(l)}_{j}).
\end{gather}
Identifying $\mathbb{E}(T_{j \pm 1}^{(k)})$ in Eq (\ref{eq:generaldiscretemoment_hetero}) with appropriate Taylor series expansions
\begin{align*}
\mathbb{E}(T_{j-1}^{(k)}) &\equiv  M_{k}^{(i)}(x) - \Delta \dfrac{\textrm{d} M_{k}^{(i)}(x)}{\textrm{d} x} + \dfrac{\Delta^2}{2} \dfrac{\textrm{d}^2 M_{k}^{(i)}(x)}{\textrm{d} x^2}\\&\quad + \mathcal{O}(\Delta^3),\\
\mathbb{E}(T_{j+1}^{(k)}) &\equiv  M_{k}^{(i+1)}(x) + \Delta \dfrac{\textrm{d} M_{k}^{(i+1)}(x)}{\textrm{d} x} + \dfrac{\Delta^2}{2} \dfrac{\textrm{d}^2 M_{k}^{(i+1)}(x)}{\textrm{d} x^2}\\&\quad + \mathcal{O}(\Delta^3),
\end{align*}
and utilising Eq (\ref{eq:order3}) gives
\begin{gather}
\label{eq:Mk_hetero_ic2_prelim}
P_{i+1}\Delta\dfrac{\textrm{d} M_{k}^{(i+1)}(x_{i})}{\textrm{d} x} - P_{i}\Delta\dfrac{\textrm{d} M_{k}^{(i)}(x_{i})}{\textrm{d} x} = \mathcal{O}(\Delta^3).
\end{gather}
Multiplying Eq (\ref{eq:Mk_hetero_ic2_prelim}) by $\Delta/\tau$ and considering the limit as $\Delta\rightarrow 0$ and $\tau\rightarrow 0$ jointly with the ratio $\Delta^{2}/\tau$ held finite leads to the interface condition:
\begin{gather}
\label{eq:Mk_hetero_ic2}
D_{i}\dfrac{\textrm{d} M_{k}^{(i)}(x_{i})}{\textrm{d} x} = D_{i+1}\dfrac{\textrm{d} M_{k}^{(i+1)}(x_{i})}{\textrm{d} x},
\end{gather}
where $i=1,\hdots,m-1$. In summary, Eqs (\ref{eq:Mk_hetero_ode})--(\ref{eq:Mk_hetero_ic1}) and (\ref{eq:Mk_hetero_ic2}) define a boundary value problem for the $k$th moment of particle lifetime in the heterogeneous environment for all $k = 1,2,3,\hdots$. Here, setting $k=1$ corresponds to the mean particle lifetime, $k=2$ corresponds to the second moment of particle lifetime, and so on.

\section{Moment expressions}

\begin{table*}[t]
{\renewcommand\arraystretch{1.4}
\begin{tabular}{|c|}
\hline
Homogeneous random walk\\
\hline
\multicolumn{1}{|c|}{Unbiased}\\
$\displaystyle M_1(x) = \frac{x(2L-x)}{2D}$\\
$\displaystyle M_2(x) = \frac{x(2L-x)}{12D^2}\left(4L^2+2Lx - x^2\right)$\\
$\displaystyle M_3(x) = \frac{x\left(2L - x \right)}{120D^3}\left(48L^4+24L^3x-8L^2x^2-4Lx^3+x^4 \right)$\\
$\displaystyle M_4(x) = \frac{x\left(2L - x \right)}{1680D^4}\left(1088L^6+544L^5x-176L^4x^2-88L^3x^3+12L^2x^4+6Lx^{5}-x^6 \right)$\\[0.2cm]
\multicolumn{1}{|c|}{Biased}\\
$\displaystyle \mathbb{M}_1(x) = \frac{1}{v^{2}}\left[De^{-Lv/D}-De^{-(L-x)v/D}+vx\right]$\\
$\displaystyle \mathbb{M}_2(x) = \frac{1}{v^{4}}\left[\left(4D^2 + 4LDv + 2D^2e^{-Lv/D}\right)\left(e^{-Lv/D}-e^{-(L-x)v/D}\right)\right.$\\
$\displaystyle\left. +2Dvx\left(e^{-Lv/D} + e^{-(L-x)v/D}\right) + v^2x^2+ 2Dvx\right]$\\[0.2cm]
\hline
Heterogeneous random walk\\
\hline
\multicolumn{1}{|c|}{2 layers}\\
$\displaystyle M_{1}^{(1)}(x) = \frac{L}{D_{1}}x - \frac{1}{2}\frac{x^{2}}{D_{1}},\quad 0 < x < x_1$\\
$\displaystyle M_{1}^{(2)}(x) = \frac{1}{2}l_{1}^{2}\left(\frac{1}{D_{1}}-\frac{1}{D_{2}}\right) + \ell_{1}\ell_{2}\left(\frac{1}{D_{1}}-\frac{1}{D_{2}}\right) + \frac{L}{D_{2}}x - \frac{1}{2}\frac{x^{2}}{D_{2}},\quad x_1 < x < x_2$\\[0.2cm]
\multicolumn{1}{|c|}{$m$ layers}\\
$\displaystyle M_{1}^{(i)}(x) =  \frac{1}{2}\sum_{k=1}^{i-1}\ell_{k}^{2}\left[\frac{1}{D_{k}} - \frac{1}{D_{i}}\right] + \sum_{k=1}^{i-1}\sum_{j=k+1}^{m}\ell_{k}\ell_{j}\left[\frac{1}{D_{k}} - \frac{1}{D_{i}}\right] + \frac{Lx}{D_{i}} - \frac{1}{2}\frac{x^{2}}{D_{i}},\quad x_{i-1} < x < x_{i}$,\\ 
$\displaystyle i = 1,\hdots,m$\\
\hline
\end{tabular}}
\caption{Moment expressions for the biased and unbiased homogeneous random walk and the unbiased heterogeneous random walk outlined in Section \ref{sec:stochastic_transport_model}. For the biased and unbiased homogeneous random walk the moments are obtained by solving the family of boundary value problems described by Eqs (\ref{eq:Mk_ode_biased})--(\ref{eq:Mk_bcs_biased}) and Eqs (\ref{eq:Mk_ode_unbiased})--(\ref{eq:Mk_bcs_unbiased}), for $k = 1,\hdots,4$ and $k = 1,2$, respectively. For the unbiased heterogeneous random walk the moments are obtained by solving the boundary value problem described by Eqs (\ref{eq:Mk_hetero_ode})--(\ref{eq:Mk_hetero_ic1}) and (\ref{eq:Mk_hetero_ic2}) with $k = 1$.}
\label{tab:1}
\end{table*}

A key benefit of working with the moments of particle lifetime is that the governing boundary value problems, derived in Section \ref{sec:stochastic_transport_model}, can be solved exactly to provide explicit information about the moments of particle lifetime. For modest $k$ and $m$, the boundary value problems can be easily solved by hand. While the algebraic details become more tedious for larger values of $k$ and $m$, the exact solutions can be computed very quickly with standard symbolic software. In Table \ref{tab:1}, we give expressions for the first four moments of particle lifetime for the unbiased homogeneous random walk and the first two moments for the biased homogeneous random walk. For the unbiased heterogeneous random walk, it is quite simple to derive by hand the expressions for the first moment in a two-layer heterogeneous medium given in Table \ref{tab:1}. Repeating this process symbolically for larger numbers of layers ($m = 3,4,\hdots$) and identifying the pattern that emerges allows the general expression given in Table \ref{tab:1} to be identified. For all other moments, Maple worksheets are provided in a GitHub repository (\href{https://github.com/elliotcarr/Carr2019a}{https://github.com/elliotcarr/Carr2019a}) for symbolically calculating the expressions.

\section{Homogenization approaches}
\label{sec:homogenization}
Consider the unbiased heterogeneous random walk and the particle lifetime distribution for an ensemble of particles released at the lattice site, $j = N$, located at the right boundary $x = L$. In this section, we explore three methods for homogenizing this stochastic process that replace the heterogeneous medium, with known diffusivities $D_{i}$ and layer widths $\ell_{i} = x_{i} - x_{i-1}$ for all $i = 1,\hdots,m$, with an equivalent or effective homogeneous medium, with effective transport coefficients.

\textit{Approximation 1}: The first approach assumes the homogenized model takes the form of an unbiased random walk with diffusivity $D_{\mathrm{eff}}$ for all $0 < x < L$. The effective diffusivity $D_{\mathrm{eff}}$ is chosen to ensure the mean particle lifetime of the homogenized model is equal to the mean particle lifetime of the heterogeneous model at the location where the particles are released ($x = L = \sum_{i=1}^{m}\ell_{i}$):
\begin{align}
\label{eq:method1_equation}
M_1^{\mathrm{eff}}(L) = M_{1}^{(m)}(L),
\end{align}
where $M_1^{\mathrm{eff}}(L)$ denotes $M_1(L)$ with $D = D_{\mathrm{eff}}$. With the expressions for $M_{1}(x)$ and $M_{1}^{(m)}(x)$ given in Table \ref{tab:1}, Eq (\ref{eq:method1_equation}) becomes:
\begin{align*}
\frac{L^{2}}{2D_{\mathrm{eff}}} = \frac{1}{2}\left(\sum_{i=1}^{m}\frac{\ell_{i}^{2}}{D_{i}} + 2\sum_{i=1}^{m-1}\sum_{j=i+1}^{m}\frac{\ell_{i}\ell_{j}}{D_{i}}\right),
\end{align*}
which can be rearranged to give the following formula for the effective diffusivity:
\begin{align}
\label{eq:Deff}
D_{\mathrm{eff}} = L^{2}\left(\sum_{i=1}^{m}\frac{\ell_{i}^{2}}{D_{i}} + 2\sum_{i=1}^{m-1}\sum_{j=i+1}^{m}\frac{\ell_{i}\ell_{j}}{D_{i}}\right)^{-1}.
\end{align}
Eq (\ref{eq:Deff}) describes how $D_{\mathrm{eff}}$ varies according to the lengths and diffusivities of the individual layers and indicates that the order in which the layers are arranged is important. Applying Eq (\ref{eq:Deff}) for a homogeneous medium, by setting either $m = 1$ and $D_{1} = D$ or $D_{i} = D$ for all $i = 1,\hdots,m$, correctly yields $D_{\mathrm{eff}} = D$.  This first homogenization approximation is widely used in the literature~\cite{Meinecke2016a,Meinecke2016b,Meinecke2017}.

\textit{Approximation 2}: The second approach also assumes the homogenized model takes the form of an unbiased random walk with diffusivity $D_{\mathrm{eff}}$ for all $0 < x < L$. However, a different approach is taken here to calculate the effective diffusivity $D_{\mathrm{eff}}$ with higher moments incorporated into the calculations. This is achieved by choosing $D_{\mathrm{eff}}$ in an attempt to ensure that the first $p$ moments of the homogenized model are equal to the first $p$ moments of the heterogeneous model at $x = L$:
\begin{align}
\label{eq:method2_overdetermined_system1}
M_1^{\mathrm{eff}}(L) &= M_{1}^{(m)}(L),\\
\label{eq:method2_overdetermined_system2}
[M_2^{\mathrm{eff}}(L)]^{1/2} &= [M_{2}^{(m)}(L)]^{1/2},\\
&\,\,\,\vdots\nonumber\\
\label{eq:method2_overdetermined_systemp}
[M_p^{\mathrm{eff}}(L)]^{1/p} &= [M_{p}^{(m)}(L)]^{1/p},
\end{align}
where $M_k^{\mathrm{eff}}(L)$ denotes $M_k(L)$ with $D = D_{\mathrm{eff}}$. To proceed, we first derive a general expression for $M_{k}(L)$, for all $k = 1,\hdots,p$. The first four moments listed in Table \ref{tab:1} give:
\begin{gather}
\label{eq:M14_unbiased_L1}
M_{1}(L) = \frac{1}{2}\frac{L^{2}}{D},\quad M_{2}(L) = \frac{5}{12}\frac{L^{4}}{D^{2}},\\
\label{eq:M14_unbiased_L2}
M_{3}(L) = \frac{61}{120}\frac{L^{6}}{D^{3}},\quad M_{4}(L) = \frac{1385}{1680}\frac{L^{8}}{D^{4}}.
\end{gather}
The numbers 1, 5, 61, 1385 appearing in the numerators in Eqs (\ref{eq:M14_unbiased_L1})--(\ref{eq:M14_unbiased_L2}) are the second to fifth Euler numbers (\href{http://oeis.org/A000364}{http://oeis.org/A000364}) while the numbers 2, 12, 120, 1680 appearing in the denominators are the second to fifth numbers in the integer sequence, $(2k)!/k!$ ($k = 0,1,\hdots$) (\href{https://oeis.org/A001813}{https://oeis.org/A001813}). These observations lead to the following general formula:
\begin{align}
\label{eq:Mk_general}
M_{k}(L) = \alpha_{k}\frac{L^{2k}}{D^{k}},\quad \alpha_{k} = \frac{k! E_{k}}{(2k)!},
\end{align}
where $E_{k}$ denotes the $k$th Euler number. Eq (\ref{eq:Mk_general}) can be easily verified symbolically for arbitrarily large $k$.

Substituting the general expression, Eq (\ref{eq:Mk_general}), into Eqs (\ref{eq:method2_overdetermined_system1})--(\ref{eq:method2_overdetermined_systemp}) and multiplying each equation through by $D_{\mathrm{eff}}$ yields an overdetermined system of $p$ linear equations in one unknown $D_{\mathrm{eff}}$:
\begin{align}
\label{eq:method2_linear_system1}
\alpha_{1}L^2 &= D_{\mathrm{eff}}M_{1}^{(m)}(L),\\
\label{eq:method2_linear_system2}
\alpha_{2}^{1/2}L^{2} &= D_{\mathrm{eff}}[M_{2}^{(m)}(L)]^{1/2},\\
&\,\,\,\vdots\nonumber\\
\label{eq:method2_linear_systemp}
\alpha_{p}^{1/p}L^{2} &= D_{\mathrm{eff}}[M_{p}^{(m)}(L)]^{1/p}.
\end{align}
Eqs (\ref{eq:method2_linear_system1})--(\ref{eq:method2_linear_systemp}) do not have an exact solution so it is impossible to match more than one moment simultaneously. A reasonable choice for $D_{\mathrm{eff}}$ is the least squares solution of Eqs (\ref{eq:method2_linear_system1})--(\ref{eq:method2_linear_systemp}). Here, $D_{\mathrm{eff}}$ satisfies the normal equations corresponding to Eqs (\ref{eq:method2_linear_system1})--(\ref{eq:method2_linear_systemp}), which take the form of a single linear equation:
\begin{align}
D_{\mathrm{eff}}\sum_{k=1}^{p}[M_{k}^{(m)}(L)]^{2/k} = L^{2}\sum_{k=1}^{p}\alpha_{k}^{1/k}[M_{k}^{(m)}(L)]^{1/k}.
\end{align}
Rearranging this equation gives the following alternative formula for the effective diffusivity:
\begin{align}
\label{eq:Deff2}
D_{\mathrm{eff}} = L^{2}\frac{\sum_{k=1}^{p}\alpha_{k}^{1/k}[M_{k}^{(m)}(L)]^{1/k}}{\sum_{k=1}^{p}[M_{k}^{(m)}(L)]^{2/k}}.
\end{align}
Eq (\ref{eq:Deff2}) provides a generalisation of Eq (\ref{eq:Deff}) that incorporates the first $p$ moments with Eq (\ref{eq:Deff2}) reducing to Eq (\ref{eq:Deff}) when $p = 1$. This can be explained by noting that Eqs (\ref{eq:method2_overdetermined_system1})--(\ref{eq:method2_overdetermined_systemp}) with $p = 1$ reduce to Eq (\ref{eq:method1_equation}), where the least squares and exact solutions are equivalent.

\textit{Approximation 3:} The third approach assumes the homogenized model takes the form of a biased random walk with diffusivity $D_{\mathrm{eff}}$ and drift $v_{\mathrm{eff}}$ for all $0 < x < L$.  Our justification for assuming the homogenized model exhibits drift stems from the lattice sites located at the interfaces between adjacent layers. At these sites the motion is biased and this bias cannot be directly accounted for solely by diffusive transport. With the homogenized model,   $D_{\mathrm{eff}}$ and $v_{\mathrm{eff}}$ are chosen to ensure the first and second moments of particle lifetime for the homogenized model are equal to the first and second moments of particle lifetime of the heterogeneous model at $x = L$:
\begin{align}
\label{eq:method3_equation1}
\mathbb{M}_1^{\mathrm{eff}}(L) = M_{1}^{(m)}(L),\\
\label{eq:method3_equation2}
\mathbb{M}_2^{\mathrm{eff}}(L) = M_{2}^{(m)}(L),
\end{align}
where $\mathbb{M}_k^{\mathrm{eff}}(L)$ denotes $\mathbb{M}_k(L)$ with $D = D_{\mathrm{eff}}$. Using the expressions for $\mathbb{M}_k(L)$ given in Table \ref{tab:1} in Eqs (\ref{eq:method3_equation1})--(\ref{eq:method3_equation2}) yields a system of two nonlinear equations in two unknowns:
\begin{widetext}
\begin{align}
\label{eq:method3_equation1a}
\frac{1}{v_{\mathrm{eff}}^{2}}\left(D_{\mathrm{eff}}e^{-Lv_{\mathrm{eff}}/D_{\mathrm{eff}}}-D_{\mathrm{eff}}+Lv_{\mathrm{eff}}\right) &= M_{1}^{(m)}(L),\\
\label{eq:method3_equation2a}
\frac{1}{v_{\mathrm{eff}}^{4}}\biggl[2D_{\mathrm{eff}}e^{-Lv_{\mathrm{eff}}/D_{\mathrm{eff}}}\left(D_{\mathrm{eff}}e^{-Lv_{\mathrm{eff}}/D_{\mathrm{eff}}} + D_{\mathrm{eff}} + 3Lv_{\mathrm{eff}}\right)-4D_{\mathrm{eff}}^2+L^2v_{\mathrm{eff}}^2\biggr] &= M_{2}^{(m)}(L),
\end{align}
\end{widetext}
where $M_{1}^{(m)}(L)$ and $M_{2}^{(m)}(L)$ are known quantities that depend on the diffusivities $D_{i}$ and layer widths $\ell_{i}$ of the heterogeneous environment. Due to the nonlinearity of Eqs (\ref{eq:method3_equation1a})--(\ref{eq:method3_equation2a}), obtaining closed-form expressions for $D_{\mathrm{eff}}$ and $v_{\mathrm{eff}}$ is not possible. Instead, we solve the system numerically. To avoid division by $v_{\mathrm{eff}}$ in Eqs (\ref{eq:method3_equation1a})--(\ref{eq:method3_equation2a}), we first expand the exponential functions in Taylor series, which produces the following equivalent system:
\begin{align}
\label{eq:method3_equation1b}
\frac{1}{2}\frac{L^2}{D_{\mathrm{eff}}} + \sum_{k=3}^{\infty}\beta_{k,1}\frac{L^{k}v_{\mathrm{eff}}^{k-2}}{k!D_{\mathrm{eff}}^{k-1}} &= M_{1}^{(m)}(L),\\
\label{eq:method3_equation2b}
\frac{5}{12}\frac{L^4}{D_{\mathrm{eff}}^2} + 2\sum_{k=5}^{\infty}\beta_{k,2}\frac{L^{k}v_{\mathrm{eff}}^{k-4}}{k!D_{\mathrm{eff}}^{k-2}} &= M_{2}^{(m)}(L),
\end{align}
where $\beta_{k,1} = (-1)^{k}$ and $\beta_{k,2} = (1-3k)(-1)^{k} + (-2)^{k}$. The left-hand sides of Eqs (\ref{eq:method3_equation1b})--(\ref{eq:method3_equation2b}) indicate that, at $x = L$, the first (second) moment of particle lifetime for the biased homogeneous random walk is given by the sum of the first (second) moment of particle lifetime for the unbiased homogeneous random walk (see Eq (\ref{eq:M14_unbiased_L1})) and a complicated correction term. For a homogeneous medium, with $m = 1$ and $D_{1} = D$ or $D_{i} = D$ for all $i = 1,\hdots,m$, $M_{1}^{(m)}(L)$ and $M_{2}^{(m)}(L)$ are given by the first and second moments in Eq (\ref{eq:M14_unbiased_L1}), respectively. In this case, the method outlined here correctly concludes that there is no drift with the solution of Eqs (\ref{eq:method3_equation1b})--(\ref{eq:method3_equation2b}) given by $D_{\mathrm{eff}} = D$ and $v_{\mathrm{eff}} = 0$.

\cbl\textit{Approximation 4:} Similarly to the first and second approaches, the fourth approach assumes the homogenized model takes the form of an unbiased random walk with diffusivity $D_{\mathrm{eff}}$ for all $0 < x < L$. Here, we use the classical homogenization result that the effective diffusivity for a layered medium is given by the harmonic average of the individual layer diffusivities in the direction normal to the layers \cite{Crank1975,Carr2017b}:
\begin{align}
\label{eq:Deff4}
D_{\mathrm{eff}} = L\left(\sum_{i=1}^{m}\frac{\ell_{i}}{D_{i}}\right)^{-1}.
\end{align}
Eq (\ref{eq:Deff4}) can be derived by using arguments presented for the continuum diffusion equation \cite{Huysmans2007} that impose that the diffusive flux at $x = L$ in the homogenized medium is equal to the diffusive flux at $x = L$ in the heterogeneous medium.
\cb

\section{Results and discussion}

\begin{figure*}[t]
\centering
\includegraphics[width=1.0\textwidth]{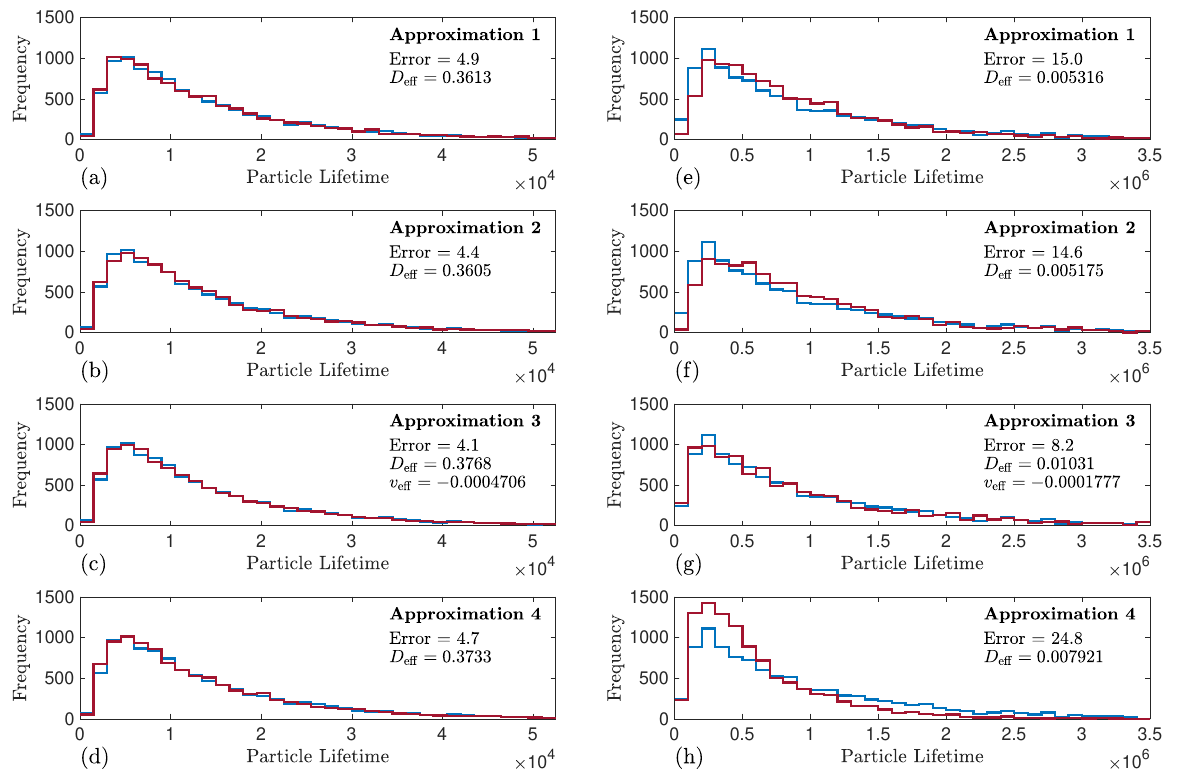}
\caption{Homogenization results for a two-layer heterogeneous system with \cbl(a)--(d)\cb{} weak and \cbl(e)--(h)\cb{} strong heterogeneity. The particle lifetime distribution for the heterogeneous (blue) and homogenized (red) systems are superimposed. For the heterogeneous system we have $[x_{0},x_{1},x_{2}] = [0,50,100]$, $[\ell_{1},\ell_{2}] = [50,50]$, $\Delta = 1$, $\tau = 1$ with $[P_{1},P_{2}] = [D_{1},D_{2}] = [0.35,0.4]$ and $[P_{1},P_{2}] = [D_{1},D_{2}] = [0.004,0.4]$ for weak and strong heterogeneity, respectively. For the homogenized random walk the parameter values are $L = 100$, $\Delta = 1$ and $\tau = 1$ with effective transport parameters shown. All particle lifetime distributions are constructed by performing 10,000 realizations of the random walk, each time releasing a particle at $x = 100$, and recording the particle's lifetime. Histograms in \cbl(a)--(d)\cb{} use bins of width 1500 while the histograms in \cbl(e)--(h)\cb{} use bins of width 100,000. The error is calculated as the Euclidean distance of the difference between the histogram bin frequencies scaled by the number of bins.}
\label{fig:2}
\end{figure*}

We now compare the accuracy of our three homogenization methods by considering the homogenization of a random walk in a heterogeneous medium containing two layers ($m = 2$) with $N = 101$, $[x_{0},x_{1},x_{2}] = [0,50,100]$, $[\ell_{1},\ell_{2}] = [50,50]$, $\Delta = 1$ and $\tau = 1$. We first consider a weakly heterogeneous medium with $[P_{1},P_{2}] = [0.35,0.4]$, and then a strongly heterogeneous medium with $[P_{1},P_{2}] = [0.004,0.4]$.  For our choice of $\Delta$ and $\tau$,  $[D_{1},D_{2}] = [P_{1},P_{2}] = [0.35,0.4]$ for the weakly heterogeneous case and $[D_{1},D_{2}] = [P_{1},P_{2}] = [0.004,0.4]$ for the strongly heterogeneous case. Performing the heterogeneous random walk 10,000 times, each time releasing a particle at $x = 100$ and recording the time taken to be absorbed at  $x = 0$, we construct a histogram of the particle lifetime, shown in Figs \cbl \ref{fig:2}(a)--(d) and \ref{fig:2}(e)--(h) \cb for the weakly and strongly heterogeneous media, respectively. Our goal is to replicate these target distributions as accurately as possible using a random walk in a homogenized medium on the same lattice with the same time step duration ($L = 100$, $\Delta = 1$, $\tau = 1$). The first step is to calculate the effective transport coefficients for the homogenized medium. For Approximation 1 and 2, $D_{\mathrm{eff}}$ is calculated directly by applying Eq (\ref{eq:Deff}) and Eq (\ref{eq:Deff2}), respectively. For Approximation 3, we solve the nonlinear system described by Eqs (\ref{eq:method3_equation1b})--(\ref{eq:method3_equation2b}) for $D_{\mathrm{eff}}$ and $v_{\mathrm{eff}}$ numerically using MATLAB's \texttt{lsqnonlin} function~\cite{Matlab2018}. The summations in Eqs (\ref{eq:method3_equation1b})--(\ref{eq:method3_equation2b}) are truncated at $k = 100$. With these computed values of the effective transport coefficients, to simulate the homogenized random walk model, the required probabilities, $P_{l}$ and $P_{r}$, are computed from Eq (\ref{eq:D_v}) with $D = D_{\mathrm{eff}}$ and $v = v_{\mathrm{eff}}$ giving $P_{l} = \tau\left[D_{\mathrm{eff}}/\Delta^2 + v_{\mathrm{eff}}/(2\Delta)\right]$ and $P_{r} = \tau\left[D_{\mathrm{eff}}/\Delta^2 - v_{\mathrm{eff}}/(2\Delta)\right]$. For Approximation 1 and 2, which assume the homogenized model takes the form of an unbiased random walk, $v_{\mathrm{eff}} = 0$ so $P_{l} = P_{r} = \tau D_{\mathrm{eff}}/\Delta^2$. Repeating the same experiment by performing the homogenized random walk 10,000 times, each time releasing a particle at $x = 100$ and recording the particle's lifetime allows histograms of the particle lifetime for each approximation method to be constructed, as shown in red in \cbl Figs \ref{fig:2}(a)--(d) and Figs \ref{fig:2}(e)--(h) \cb for the weakly and strongly heterogeneous cases, respectively. To quantify the error of the homogenized distribution compared to the target distribution we calculate the Euclidean distance of the difference between the histogram bin frequencies scaled by the number of bins (denoted by Error in Fig \ref{fig:2}).

For the weakly heterogeneous test case we have $D_{\mathrm{eff}} = 0.3613$  for Approximation 1, and $D_{\mathrm{eff}} = 0.3605$ for Approximation 2 with $p = 2$.  Therefore, in this case, defining the homogenized medium by matching the mean first exit time (Approximation 1) gives very similar results than if we homogenize by matching the first two moments of the exit time distribution (in a least squares sense).  This is reassuring as Approximation 1 is widely invoked but the question of whether the homogenized media also captures higher moments of particle lifetime is never explicitly tested~\cite{Meinecke2016a,Meinecke2016b,Meinecke2017}. \cbl Approximation 4 yields $D_{\mathrm{eff}} = 0.3733$ for the weakly heterogeneous test case and gives very similar results to the first two approximations. \cb For the weakly heterogeneous test case Approximation 3 gives $D_{\mathrm{eff}} = 0.3768$ and $v_{\mathrm{eff}} = -0.0004706$. Here, we have a very small negative effective drift, implying a particle in the homogenized system is slightly biased towards transport in the positive $x$ direction, see Eq (\ref{eq:D_v}), which makes sense given that a particle at the interface ($x = 50$) is more likely to take one step in the positive $x$ direction than in the negative $x$ direction since $P_{2} = 0.4 > P_{1} = 0.35$. Overall, for the weakly heterogeneous test case we see that all \cbl four \cb approximations give very similar results, and we have an excellent match between the homogenized lifetime distributions and the target lifetime distribution in Figs \cbl \ref{fig:2}(a)--(d)\cb.

The situation is different for the strongly heterogeneous medium where $D_{\mathrm{eff}} = 0.005316$  for Approximation 1, $D_{\mathrm{eff}} = 0.005175$ for Approximation 2 with $p = 2$, $D_{\mathrm{eff}} = 0.01031$ and $v_{\mathrm{eff}} = -0.0001777$ for Approximation 3 \cbl and $D_{\mathrm{eff}} = 0.007921$ for Approximation 4. \cb The ratio $v_{\mathrm{eff}}\Delta/D_{\mathrm{eff}}$ under Approximation 3 is an order of magnitude larger for the strongly heterogeneous medium compared to the weakly heterogeneous medium indicating that drift is more dominant under strong heterogeneity. This observation is evident in Fig \cbl \ref{fig:2}(e)--(g) \cb, with Approximation 3 clearly producing a better match with the target distribution, particularly for small particle lifetimes. \cbl Approximation 4 overestimates the effective diffusivity, which leads to a gross overestimation of the proportion of smaller particle lifetimes as evident in Fig \ref{fig:2}(h). \cb Across both test cases, when the homogenized model is assumed to take the form of an unbiased random walk, we observe little benefit of including higher moments in the calculation of $D_{\mathrm{eff}}$ with Approximation 1 and 2, the latter with $p = 2$, providing an equivalent level of accuracy. Similar observations are observed for larger values of $p$.  Therefore, caution is warranted when using homogenization approximations to deal with strongly heterogeneous media, as can be the case in biological environments~\cite{Swanson2000}, since the values of the effective transport coefficients can depend upon the homogenization approximation and it is not always clear which approximation will provide the best result. \cbl When strong differences in transport coefficients are present, the various approximations we study here can give different effective transport coefficients because each homgenization approximation relies on matching different properties between the heterogeneous system and the effective homogeneous system. \cb

\section{Conclusions}
In this work we construct new homogenization approximations allowing us to approximate a diffusive transport process, on an arbitrarily heterogeneous one-dimensional domain, using an equivalent homogenized medium.  This kind of homogenization approximation is often used to circumvent issues associated with negative transition rates that can arise when standard transport operators are discretized on heterogeneous domains, and the most common approach is to replace the heterogeneous medium with an homogenized medium so that the mean particle lifetime is preserved. Here we take a more general approach and explain how to derive exact closed form expressions for the $k$th moment of particle lifetime in an arbitrary heterogeneous system consisting of $m$ distinct layers.  We explain how to arrive at a family of boundary value problems from an underlying random walk process, taking great care to explore how the interface conditions in the discrete random walk formulation translate into boundary conditions in the continuous description.

Given closed-form expressions for the $k$th moment of particle lifetime for an arbitrarily heterogeneous domain we can define several approximations from which we can construct effective transport parameters in a homogenized model.  We use three approximations: first we match the mean particle lifetime; second we match the first $p>1$ moments of particle lifetime in a least squares sense; and third we match the particle lifetime distribution from the heterogeneous system with a homogenized advection-diffusion model where the different jump rates in the various layers give rise to a net bias, leading to non-zero advection, for the entire homogenized system.  Example calculations show that all three approximations lead to similar results in the limit of weak heterogeneity whereas the three approximations can lead to very different estimates of the effective transport coefficients in the case of strong heterogeneity.  This result suggests that care needs to be exercised when homogenizing strongly heterogeneous environments.  In this case, it is prudent to apply multiple homogenization criteria, such as applying Approximation 2 with various values of $p$, to provide an understanding of the sensitivity of the results to the homogenization criteria.  Our approach, which leads to exact closed-form expressions for the $k$th moment of particle lifetime in an arbitrarily heterogeneous system can be easily used to test various homogenization criteria, as we demonstrate in Fig \ref{fig:2}.

The results in this study can be extended in many ways.  For simplicity, here we focus on one-dimensional Cartesian domains.  However, the concepts outlined here generalize to higher dimensional problems, such domains with radial and spherical symmetry where exact solutions for the $k$th moment of particle lifetime can also be calculated exactly using symbolic software~\cite{Simpson2015}.  In contrast, for higher-dimensional problems without radial or spherical symmetry, the same ideas pursued here can be used to construct elliptic boundary value problems that govern the $k$th moment of particle lifetime, however the solution of these boundary value problems would be more easily found numerically.

\noindent
\textit{Acknowledgements}.\ This work is supported by the Australian Research Council (DE150101137, DP170100474). \cbl We appreciate helpful comments and suggestions from the referee.\cb


\end{document}